\documentclass[12pt]{iopart}

\usepackage{graphicx}
\usepackage{dcolumn}
\usepackage{bm}
\usepackage{hyperref}

\usepackage{multirow}%
\usepackage{amsmath,amssymb,amsfonts}%
\usepackage{tabularx,booktabs}
\usepackage{mathtools}

\newcommand{\KN}{\mathbin{\bigcirc\mspace{-15mu}\wedge\mspace{3mu}}}

\begin{document}

\title[Encoding the Einstein Equations into an Algebraic Commutator Condition]{Encoding the Einstein Equations into an Algebraic Commutator Condition}

\author{Jack C. M. Hughes \& Fedor V. Kusmartsev}

\address{College of Engineering and Physical Sciences, Khalifa University, PO Box 127788, Abu Dhabi, United Arab Emirates}
\ead{jack.hughes@ku.ac.ae}
\vspace{10pt}
\begin{indented}
\item[]November 2025 
\end{indented}

\begin{abstract}
We show that the structure of the Lorentz group in four dimensions is such that unimodular (trace-free) gravity can be consistently represented as an algebraic condition on the symmetric product space of 2-forms. This condition states that the commutator between the Riemann tensor and the Hodge dual must be equal to the commutator between the Kulkarni-Nomizu product of the energy-momentum and the metric with the Hodge dual; symbolically, $[\text{Riem}, \star] = 4\pi [T\KN g, \star]$. We show that this condition is equivalent to the trace-free field equations, that the right-hand-side vanishes if and only if the energy-momentum tensor vanishes (recovering the appropriate Einstein spacetime limit) and that this condition can be solved for electrovacuum in the spherically symmetric ansatz to yield Reissner-Nordstr\"om-de Sitter uniquely. This analysis suggests that the conceptual distinction between unimodular gravity and General Relativity is one of emphasis on how irreducible representations of the Riemann tensor are constrained by the existence of energy-momentum and the associated field equations. 
\end{abstract}

\section{Introduction}
Depending on what dimension for spacetime is chosen, General Relativity (GR) will have properties unique to that dimension. In three dimensions for instance, vacuum GR (with or without the cosmological constant $\Lambda$) can be expressed as a topological gauge theory \cite{Witten1988, AchucarroTownsend1986, deBoerJottar2014, Wise2009SIGMA080}. In four dimensions however there exists unique algebraic structures as a consequence of Hodge duality \cite{Besse1987, Krasnov2011}. This can be understood at various levels, the most fundamental of which is via the semi-simple structure of the (complexified) Lorentz algebra \cite{Krasnov2011, Harvey1990SpinorsCalibrations}: 
\begin{equation}\label{Lie_Alg_Decomp}
    \mathfrak{so}(1,3)_{\mathbb{C}} \cong \mathfrak{sl}(2,\mathbb{C}) \oplus \overline{\mathfrak{sl}(2,\mathbb{C})}.
\end{equation}
Many of the tensor spaces of interest in GR utilize this splitting in the following sense: labeling the irreducible representations of the Lorentz algebra (\ref{Lie_Alg_Decomp}) as $(j_1, j_2)$ \cite{Srednicki2007}, the spaces
\begin{equation}
    \Omega(j) = (j, 0) \oplus (0, j), \quad \Sigma(j) = (j, j), \quad j= 0, \frac{1}{2}, 1, \cdots 
\end{equation}
are of importance in the irreducible decomposition of various natural tensor fields. It is the Hodge dual that is responsible for connecting representation theory to geometric structure, since $\Omega(1)$ and $\Omega(2)$ - corresponding to the 2-forms and Weyl tensors\footnote{In the sense that any tensor field in this representation has the same symmetries as the Weyl tensor, together with vanishing trace.} respectively - are projections into the complex eigenspaces of the Hodge dual, which squares to minus the identity on 2-forms \cite{Besse1987, Krasnov2011}. This allows for some strong statements to be deduced regarding the curvature of spacetime, which i) being a 4-tensor with various symmetries and satisfying the first Bianchi identity has its standard orthogonal (Ricci) decomposition into Weyl, trace-free Ricci and scalar components \cite{Besse1987, Besse1981, BergerGauduchonMazet1971, Fre2013GravityVol1, Weinberg1972}, and ii) being also a 2-form has its Hodge decomposition into complex eigenspaces via (\ref{Lie_Alg_Decomp}) \cite{Besse1987, Krasnov2020, Plebanski1977Separation, ChungHwangYang2023}. One finds that in four dimensions \textit{only} the Riemann tensor is an element of the space \cite{Besse1987}
\begin{equation}\label{Irr_Rep_Rie}
    \begin{split}
        \text{Riem} & \in  \Omega(2) \oplus \Sigma(1) \oplus \Sigma(0) = (2,0) \oplus (0,2) \oplus (1,1) \oplus (0,0).
    \end{split}
\end{equation}
This has important consequences on the geometry and topology of such spacetimes. Indeed, if in addition the manifold is restricted to be \textit{Einstein} (meaning that the Ricci tensor is proportional to the metric)
\begin{equation}\label{EC}
    R_{\mu \nu} \propto g_{\mu \nu},
\end{equation}
then the duality structure (\ref{Lie_Alg_Decomp}) restricts the possible irreducible representations of the curvature tensor (\ref{Irr_Rep_Rie}) \cite{Besse1987, Krasnov2020, Plebanski1977Separation, Atiyah1978, Park2022AnatomyEinstein}. In this dimension the moduli space of Einstein structures is known to be highly constrained, which in many compact cases consists of only a single point \cite{Besse1987, HoKimYang2025JGP}. The rigidity is (in part) a consequence of the many ways the Einstein condition (\ref{EC}) manifests itself through the lens of (\ref{Irr_Rep_Rie}). For instance, if (\ref{EC}) is satisfied then the Levi-Civita connection must be self-dual \cite{Atiyah1978, SingerThorpe1975}. It is clear that this condition (and any other equivalent characterization) must hold on any solution to the vacuum Einstein field equations
\begin{equation}\label{Field_Equations}
    R_{\mu \nu} - \frac{1}{2} Rg_{\mu \nu} + \Lambda g_{\mu \nu} = 0,
\end{equation}
with constant of proportionality in (\ref{EC}) now given by the cosmological constant,
\begin{equation}\label{EC_Spacetime}
    R_{\mu \nu} = \Lambda g_{\mu \nu}.
\end{equation}
The duality structure of the gravitational vacuum in four dimensions has been of interest, especially when attempting to reformulate GR in terms of new variables \cite{Krasnov2011, Krasnov2020, Plebanski1977Separation, Krasnov2017SelfDualGravity, Mielke2017, EguchiGilkeyHanson1980, BlagojevicHehl1995, NakamichiSugamotoOda1991, SmolinStarodubtsev2003, DePietriFreidel1999, CapovillaDellJacobsonMason1991, Giulini1994, Urbantke1984, Ward1980SelfDualLambda, AlexandrovPiolineVandoren2010, NietoObregonSocorro1994, Robinson1995LagrangianEYM, Robinson1994GL2C_EinsteinMaxwell, Carmeli1972SL2C, Nissani1984SL2CGauge, IshibashiSpeziale2009, tHooft1991ChiralAlternative, LiuMontesinosPerez2010, Reisenberger1997LeftHandedSimplicial, Smolin2009PlebanskiUnification, SmolinSpeziale2010PlebanskiCosmoImmirzi, Zapata1996TopologicalLattice}. The key observation is that four-dimensional GR in vacuum requires only the self-dual sector of curvature, so one can project the Einstein-Cartan action into this eigenbasis of the Hodge dual with no loss in information \cite{Krasnov2020, Atiyah1978, JacobsonSmolin1987}. The Weyl spinor representations $(1/2, 0)$ then generate this sector, since \cite{Srednicki2007}
\begin{equation}
    (1/2,0) \otimes (1/2,0) \cong (0,0) \oplus (1,0),
\end{equation}
giving a spinor basis for the self-dual 2-forms consistent with the Plebanski formalism \cite{Krasnov2020, CapovillaDellJacobsonMason1991, Samuel1987, JacobsonSmolin1988, Samuel1992}. This is the algebra underlying the Petrov classification \cite{Besse1987, Plebanski1977Separation, Petrov1969, Petrov2000Classification}. Complexifying instead the underlying spacetime, the groups at play in these dual structures lift to those relevant in the Twistor construction \cite{Krasnov2020, AlexandrovPiolineVandoren2010}. Furthermore since via the theory of characteristic classes various topological invariants can be constructed as integral classes in the Riemann curvature \cite{Besse1987, Mielke2017, Tu2017}, the duality structure of such invariants in four dimensions gives a unique classification scheme of (closed) Einstein manifolds, forming a link to Yang-Mills instantons \cite{Park2022AnatomyEinstein, HoKimYang2025JGP}. The interplay between duality and topology in four dimensions also manifests itself in horizon thermodynamics \cite{HughesKusmartsev2025TopologicalOrigin, HughesKusmartsev2025TopologicalOriginPZ, Volovik2025_Top, Volovik2025Comment, HughesKusmartsev2025ReplyJETPLett, HughesKusmartsev2025ReplyPZ}.

When the Einstein condition (\ref{EC}) is satisfied in a four-dimensional spacetime, the $(1,1)$ representation in (\ref{Irr_Rep_Rie}) - which corresponds to the trace-free Ricci tensor - must vanish \cite{Besse1987, Krasnov2020}. This can be written as a compact statement using tensor indices. If $R_{\mu \nu}^{\quad \rho \sigma}$ is the Riemann tensor and $\varepsilon_{\mu \nu}^{\quad \rho \sigma}$ is the Levi-Civita tensor (what is effectively the Hodge dual on 2-forms), then a four-dimensional manifold is Einstein \textit{if and only if} the Riemann tensor commutes with the Hodge dual as (endomorphism) operators on the 2-forms \cite{Krasnov2020}:
\begin{equation}\label{Commutator}
    R_{\mu \nu}^{\quad \rho \sigma} \varepsilon_{\rho \sigma}^{\quad \alpha \beta} - \varepsilon_{\mu \nu}^{\quad \rho \sigma} R_{\rho \sigma}^{\quad \alpha \beta} = 0 \; \Leftrightarrow \; R_{\mu \nu} \propto g_{\mu \nu}.
\end{equation}
Again, any solution to the vacuum field equations (\ref{Field_Equations}) necessarily satisfies this condition in four dimensions. However as soon as energy-momentum content is present in the form of non-vanishing $T_{\mu \nu}$, (\ref{Commutator}) is no longer satisfied. Equivalently, the trace-free Ricci sector in (\ref{Irr_Rep_Rie}) does not vanish. Hence, the operator 
\begin{equation}\label{Delta}
    \Delta_{\mu \nu}^{\quad \alpha \beta} = \frac{1}{2} R_{\mu \nu}^{\quad \rho \sigma} \varepsilon_{\rho \sigma}^{\quad \alpha \beta} - \frac{1}{2}\varepsilon_{\mu \nu}^{\quad \rho \sigma} R_{\rho \sigma}^{\quad \alpha \beta} 
\end{equation}
is a \textit{measure of the degree to which a spacetime fails to be vacuum}. It is a direct encoding of how matter interacts with the duality structure of four-dimensional geometry, something that is historically non-trivial to quantify (e.g. in Plebanski theories) \cite{Krasnov2017SelfDualGravity, CapovillaDellJacobsonMason1991, CeladaGonzalezMontesinos2016}. The question we are interested in for this work is how the condition (\ref{Delta}) can be generalized to spacetimes containing arbitrary energy-momentum tensor. That is, is there a way to lift energy-momentum content to the (symmetric) product of the 2-form sector, such that gravity can be algebraically encoded as an equation within the Hodge decomposed space (\ref{Irr_Rep_Rie})? Such an algebraic description of spacetime-matter coupling is of interest from the geometrodynamics \cite{Mielke2017} perspective, being conceptually analogous to the local Rainich conditions \cite{Rainich1925ElectrodynamicsGR, MisnerWheeler1957}. 

In what follows, we demonstrate that this is indeed possible, but only up to the trace-free sector of the geometry (which leaves the scalar curvature undetermined). That is, unimodular gravity \cite{Ellis2014TraceFreeInflation, CarballoRubioGarayGarciaMoreno2023, BengocheaLeonPerezSudarsky2023} may be consistently encoded on (\ref{Irr_Rep_Rie}) through a generalization of (\ref{Commutator}) to the following (symbolic) form
\begin{equation}\label{Intro_Main}
    [\text{Riem}, \star] = 4\pi [g\KN T, \star],
\end{equation}
where $g \KN T$ is the \textit{Kulkarni–Nomizu product} between the metric and the energy-momentum tensor (see equation \ref{KN_Prod}) \cite{Besse1987}. This product operation lifts the energy-momentum tensor to the space (\ref{Irr_Rep_Rie}). In section 5, it is shown that this condition is equivalent to imposing the trace-free field equations of unimodular gravity, which together with the assumption of the conservation of energy-momentum $\nabla_{\mu} T^{\mu \nu} = 0$ is equivalent to GR \cite{Ellis2014TraceFreeInflation, CarballoRubioGarayGarciaMoreno2023, BengocheaLeonPerezSudarsky2023}. 

However it is possible to take (\ref{Intro_Main}) as the equations defining the theory, in that they may be solved in this form - or in tensor indices equation (\ref{Main_Eqn_Indices}) - without the need to project down to the trace-free equations. In section 4, it is shown using the spherically symmetric ansatz (\ref{ansatz}) and spherical potential $A = -(Q/r) dr$ that (\ref{Intro_Main}) can be solved to yield Reissner-Nordstr\"om-de Sitter uniquely. 

Taken together these arguments indicate that the trace-free theory can actually be constructed as algebraic conditions on the irreducible representations of the Riemann tensor, with the energy-momentum tensor quantifying to what extent geometry mixes the self-dual and anti-self-dual eigenspaces of the Hodge decomposition. By shifting focus away from the symmetric 2-tensor field sector of the geometry, the field equations are observed to already exist at a higher-level, with the standard (trace-free) field equations being a projection from a more representation-theoretic formulation. 

\section{Background}
In any dimension, the Riemann tensor resides within a unique structure known as \textit{the algebraic space of curvature tensors} $\mathcal{K}$ \cite{Besse1987, GallotHulinLafontaine2004}. To isolate this space, we first recognize that $R_{\mu \nu \rho \sigma}$ is an element of the fourth-order tensor product of the (co)\footnote{Algebraically there is no need to distinguish between the tangent and co-tangent bundles because the metric establishes a canonical isomorphism between the two.}tangent bundle $\otimes^4 T^*\mathcal{M}$. The (special) orthogonal group acts naturally on this space, and it is necessary to construct the irreducible representations under this action. As an example, consider the space of 2-tensors $\otimes^2 T^*\mathcal{M}$: any 2-tensor can be decomposed into its anti-symmetric ($\Lambda^2$), its symmetric trace-free ($S^2_0$) and pure-trace ($\mathbb{R}g$) components \cite{Besse1987, Srednicki2007}
\begin{equation}\label{U_Decomp}
    U \in \otimes^2 T^*\mathcal{M}, \quad U_{\mu \nu} = \underbrace{\frac{1}{2}(U_{\mu \nu} - U_{\nu \mu})}_{\Lambda^2(T^*\mathcal{M})} + \underbrace{\bigg[\frac{1}{2} (U_{\mu \nu} + U_{\nu \mu}) - \frac{g^{\alpha \beta}U_{\alpha \beta}}{n}g_{\mu\nu}\bigg]}_{S^2_0(T^*\mathcal{M})}  + \underbrace{\frac{g^{\alpha \beta}U_{\alpha \beta}}{n} g_{\mu \nu}}_{\mathbb{R}g},
\end{equation}
where in what follows an under-brace will denote the algebraic space into which a given representation falls into or coincides with. For four dimensional spacetime specifically, the standard labeling of the Lie algebra of the Lorentz group (\ref{Lie_Alg_Decomp}) reproduces (\ref{U_Decomp}) through Clebsch-Gordon decomposition \cite{Srednicki2007}:
\begin{equation}\label{U_4d}
    (\frac{1}{2}, \frac{1}{2}) \otimes (\frac{1}{2}, \frac{1}{2}) = \big[\underbrace{(1,0) \oplus (0,1)}_{{\Lambda^2 (T^*\mathcal{M})}} \big] \oplus \underbrace{(1,1)}_{S_0^2(T^*\mathcal{M})} \oplus \underbrace{(0,0)}_{\mathbb{R}g}.
\end{equation}
The chiral structure of four dimensional geometries is already evident at the level of the 2-forms. Operationally, this is reflected in the Hodge dual which acts as endomorphism on $\Lambda^2$ \cite{Krasnov2020}
\begin{equation}\label{Hodge_Op}
    \begin{split}
        \star\colon \Lambda^2 &\to \Lambda^2, \\
        \omega_{\mu \nu} & \mapsto (\star\omega)_{\mu \nu} = \frac{1}{2} \varepsilon_{\mu \nu}^{\quad \rho \sigma} \omega_{\rho \sigma},
    \end{split}
\end{equation}
where $\varepsilon_{\mu \nu}^{\quad \rho \sigma}$ is the Levi-Civita tensor \cite{Krasnov2020, Atiyah1978}. Since $\star$ squares to minus the identity on $\Lambda^2$, 
\begin{equation}\label{Hodge_Square_Prop}
    (\star^2 \omega)_{\mu \nu} = \frac{1}{4} \varepsilon_{\mu \nu}^{\quad \rho \sigma} \varepsilon_{\rho \sigma}^{\quad \alpha \beta} \omega_{\alpha \beta} =  -\delta_{[\mu}^{\alpha} \delta^{\beta}_{\nu]} \omega_{\alpha \beta}.
\end{equation}
the chiral splitting in (\ref{U_4d}) is into the \textit{complex} self-dual $\Lambda^+$ and anti-self-dual $\Lambda^-$ eigenspaces of the Hodge dual
\begin{equation}\label{Hodge_Decomp}
    \Lambda^2 = \underbrace{\Lambda^+}_{(1,0)} \oplus \underbrace{\Lambda^-}_{(0,1)},
\end{equation}
which is a projection of the fundamental semi-simple structure (\ref{Lie_Alg_Decomp}). Note that this complexification does not occur in the Euclidean or split signature cases, where the eigenspaces of the Hodge dual remain real \cite{Harvey1990SpinorsCalibrations, Zapata1996TopologicalLattice, Woit2017QuantumGroupsReps, Woit2021EuclideanTwistor}. 

To construct the algebraic space of curvature tensors $\mathcal{K}$, a similar decomposition to (\ref{U_Decomp}) for the tangent bundle product $\otimes^4 T^*\mathcal{M}$ is required. However, this space is very large (scaling with manifold dimension $n$ as $n^4$), and a number of elements of this space are redundant once the symmetries of the Riemann tensor are imposed \cite{Weinberg1972}. Provided the connection is Levi-Civita, these amount to the following index symmetries 
\begin{equation}\label{Rie_Symm}
    R_{\mu \nu \rho \sigma} = - R_{\nu \mu \rho \sigma} = -R_{\mu \nu \sigma \rho} = R_{\rho \sigma \mu \nu}.
\end{equation}
In words, this means that the Riemann tensor transforms as a 2-form in its two pairs of indices $(\mu \nu)$ and $(\rho \sigma)$, while simultaneously it is a symmetric tensor under the exchange of these pairs $(\mu \nu \leftrightarrow \rho \sigma)$.  Algebraically then, we isolate the subspace of $\otimes^4 T \mathcal{M}$ corresponding to the symmetrization $S^2$ over 2-forms $\Lambda^2$:
\begin{equation}
    S^2(\Lambda^2 \; T^*\mathcal{M}) \subset \otimes^4 T^* \mathcal{M}.
\end{equation}
This subspace has an irreducible decomposition into four sectors and is easily accessed via the Bianchi map \cite{Besse1987, GallotHulinLafontaine2004}:
\begin{equation}\label{Bianchi_Map}
    \begin{split}
        b\colon \otimes^4 T\mathcal{M} & \to S^2(\Lambda^2 T\mathcal{M}) \\
        U_{\mu \nu \rho \sigma} & \mapsto b(U)_{\mu \nu \rho \sigma} = \frac{1}{3} \big(U_{\mu \nu \rho \sigma} + U_{\nu \rho \mu \sigma} + U_{\rho \mu \nu \sigma}\big).
    \end{split}
\end{equation}
Since $b$ is clearly invariant under diffeomorphisms and acts as a projection operator ($b^2 = b$), the space $S^2(\Lambda^2 \; T^*\mathcal{M})$ admits an orthogonal decomposition into the kernel and image of the Bianchi map,
\begin{equation}
    S^2(\Lambda^2 \; T^*\mathcal{M}) = \text{im} \;b \oplus \ker b.
\end{equation}
Again in four-dimensional spacetime this result is reflected simply in Clebsch-Gordon decomposition:
\begin{equation}\label{Symmetric_Decomp}
    S^2(\Lambda^2 \; T^*\mathcal{M})  = \underbrace{\big[(2,0) \oplus (0,2)\big] \oplus (1,1) \oplus (0,0)_+}_{\text{ker}\; b} \oplus \underbrace{(0,0)_-}_{\text{im}\; b}, 
\end{equation}
where $(0,0)_{\pm}$ are the self-dual (respectively anti-self-dual) scalars induced by the symmetrization procedure: they correspond (respectively) to the two possible parities (even and odd) of the invariant bilinear forms in $\mathfrak{so}(1,3)$ \cite{Krasnov2020}:
\begin{equation}
    (0,0)_+ \cong  g_{\mu \rho}g_{\nu \sigma} - g_{\mu \sigma}g_{\nu \rho}, \quad (0,0)_-  \cong \varepsilon_{\mu \nu \rho \sigma}.
\end{equation}
$S^2(\Lambda^2 \; T^*\mathcal{M})$ is the space in which the curvature tensor resides for those theories containing spacetime torsion \cite{Mielke2017, Hehl1973, Hehl1974}. Torsion introduces a completely antisymmetric 4-form component to the curvature, which in four dimensions is proportional (up to a function) to the volume of spacetime \cite{Mielke2017, AldrovandiPereira2012}. However, if the first Bianchi identity is satisfied
\begin{equation}\label{First_Bianchi_Identity}
    R_{\mu \nu \rho \sigma} + R_{\nu \rho \mu \sigma} + R_{\rho \mu \nu \sigma} = 0,
\end{equation}
or what can be rewritten compactly in four dimensions as 
\begin{equation}
    R_{\mu \nu \rho \sigma} \varepsilon^{\mu \nu \rho \sigma} = 0,
\end{equation}
then the Riemann tensor resides only in the \textit{kernel} of the map (\ref{Bianchi_Map}), which defines the algebraic space of curvature tensors $\mathcal{K} \cong \ker b$ \cite{Besse1987, GallotHulinLafontaine2004}. In order to further classify the irreducible structure of $\mathcal{K}$, it proves necessary to ask how elements of the decomposition (\ref{U_Decomp}) may be lifted to the space $\mathcal{K}$.

\subsection{The Kulkarni–Nomizu Product}
A natural question to be asked is if a product operation can be constructed such that a combination of two 2-tensors maps into $\mathcal{K}$. That there exists such a \textit{unique} symmetric bi-linear diffeomorphism invariant operator was first shown in the works of Kulkarni \cite{Kulkarni1972} and Nomizu \cite{Nomizu1972}, provided that the individual 2-tensors are symmetric. Indeed, the Kulkarni-Nomizu (KN) product $\KN$ is given as follows \cite{Besse1987, GallotHulinLafontaine2004, AndrewsHopper2010}:
\begin{equation}\label{KN_Prod}
    \begin{split}
        \KN\colon S^2(T^*\mathcal{M}) \otimes S^2(T^*\mathcal{M}) & \to \mathcal{K}, \\
        A_{\mu \nu},B_{\mu \nu}\in S^2(T^*\mathcal{M}) & \mapsto (A \KN B)_{\mu \nu \rho \sigma} = A_{\mu \rho} B_{\nu \sigma} + A_{\nu \sigma} B_{\mu \rho} - A_{\mu \sigma} B_{\nu \rho} - A_{\nu \rho} B_{\mu \sigma}.
    \end{split}
\end{equation}
It is easily verified that $(A\KN B)_{\mu \nu \rho \sigma}$ satisfies the symmetry properties (\ref{Rie_Symm}), together with the first Bianchi identity (\ref{First_Bianchi_Identity}).

By utilizing the above, we can express elements of the irreducible decomposition of $\mathcal{K}$ in terms of KN products. For instance, taking the KN product of the metric with itself yields 
\begin{equation}\label{KN_Prop_Metric}
    (g\KN g)_{\mu \nu \rho \sigma} = g_{\mu[\rho} g_{\sigma] \nu},
\end{equation}
which amounts to the identity operator on $\Lambda^2$ (\ref{Hodge_Square_Prop}). Similarly, if $L$ is a trace-free symmetric 2-tensor - $L \in S^2_0 (T^*\mathcal{M})$ - we have
\begin{equation}
    (g \KN L)_{\mu \nu \rho \sigma} = \frac{1}{2}\big( g_{\mu [\rho} L_{\sigma] \nu} - g_{\nu [\rho} L_{\sigma] \mu}\big).
\end{equation}
With this, it can be shown that the algebraic space of curvature tensors $\mathcal{K}$ has the following fundamental irreducible decomposition into Weyl, trace-free symmetric and scalar curvature sectors \cite{Besse1987}:
\begin{equation}\label{Full_Decomp_Alg}
    \mathcal{K} \quad = \quad \mathcal{W} \quad \oplus \quad  S_0^2(T^*\mathcal{M}) \KN g \quad \oplus \quad \mathbb{R} g \KN g,
\end{equation}
which at the tensorial level is the well-known Ricci/orthogonal decomposition
\begin{equation}\label{Ricci_Decomp}
    R_{\mu \nu \rho \sigma} = \underbrace{C_{\mu \nu \rho \sigma}}_{\mathcal{W}} +  \underbrace{\frac{1}{(n-2)}\big( g_{\mu [\rho} Z_{\sigma] \nu} - g_{\nu [\rho} Z_{\sigma] \mu}\big)}_{S_0^2(T^*\mathcal{M}) \KN g} + \underbrace{\frac{2}{n(n-1)} Rg_{\mu [\rho} g_{\sigma] \nu}}_{\mathbb{R} g \KN g},
\end{equation}
where
\begin{equation}\label{Trace_Free_Ricci}
    Z_{\mu \nu} = R_{\mu \nu} - \frac{1}{n}R g_{\mu\nu}.
\end{equation}
Intuitively, the ``space of Weyl tensors" $\mathcal{W}$ is the irreducible component of $S^2(\Lambda^2 T\mathcal{M})$ orthogonal to the KN product,
\begin{equation}
    \langle \text{Weyl}, A\KN B \rangle = 0 \quad \forall A, B \in S^2(T\mathcal{M}).
\end{equation}
More abstractly, $\mathcal{W}$ is analogous to an \textit{entangled state} component of algebraic curvature tensors, while the remaining elements of (\ref{Full_Decomp_Alg}) are analogous to a separable product of pure states. Defining 
\begin{equation}\label{P_g}
    \mathcal{P}_g \coloneq \text{span}\{g \KN L\colon L \in S^2(T^*\mathcal{M})\},
\end{equation}
we have the orthogonal decomposition
\begin{equation}
    \mathcal{K} \quad = \quad \mathcal{W} \quad \oplus \quad \mathcal{P}_g.
\end{equation}
Focusing on the Riemann tensor itself, this splitting is the conceptual origin of the Schouten tensor $P$ \cite{StephaniKramerMacCallumHoenselaersHerlt2003, GriffithsPodolsky2009, ValienteKroon2016}: it is the symmetric 2-tensor that covers $\mathcal{P}_g$ upon taking the KN product with the metric,
\begin{equation}
    P_{\mu \nu} = \frac{1}{n-2} \bigg(R_{\mu \nu} - \frac{R}{2(n-1)} g_{\mu \nu} \bigg).
\end{equation}

\subsection{Duality in Four Dimensions}
A unique property of four dimensions is that the Weyl sector becomes reducible via the projection of antisymmetric indices into the eigenspaces of the Hodge dual (\ref{Hodge_Decomp}),
\begin{equation}
    \mathcal{K} \quad = \quad  \underbrace{\big[(2,0) \oplus (0,2)\big]}_{\mathcal{W}}  \quad \oplus \underbrace{(1,1)}_{S_0^2(T^*\mathcal{M}) \KN g} \oplus  \quad \underbrace{(0,0)_+}_{\mathbb{R} g \KN g}.
\end{equation}
As a consequence of the symmetries satisfied by the Riemann tensor (\ref{Rie_Symm}), it may be viewed (like the Hodge dual) as an operator on the 2-forms
\begin{equation}
    \begin{split}
        \text{Riem}\colon \Lambda^2 &\to \Lambda^2, \\
        \omega_{\mu \nu} & \mapsto (\text{Riem} \;\omega)_{\mu \nu} =  R_{\mu \nu}^{\quad \rho \sigma} \omega_{\rho \sigma}.
    \end{split}
\end{equation}
Using the Hodge decomposition (\ref{Hodge_Decomp}), we may consider instead a (symmetric) matrix representation for the Riemann tensor on the eigenbases of the Hodge dual. Explicitly, in four dimensions we may expand \cite{Krasnov2020, Besse1987, Krasnov2017SelfDualGravity}
\begin{equation}\label{SD_Decomp}
    S^2(\Lambda^2 \; T^*\mathcal{M}) = \underbrace{\big(\Lambda^+ \otimes \Lambda^+\big)}_{A} \oplus \underbrace{\big(\Lambda^+ \otimes \Lambda^-\big)}_{B} \oplus \underbrace{\big(\Lambda^- \otimes \Lambda^+\big)}_{B^T = \overline{B}} \oplus \underbrace{\big(\Lambda^- \otimes \Lambda^-\big)}_{C = \overline{A}},
\end{equation}
to give the following $6\times 6$ (with $3\times 3$ irreducible blocks) decomposition of the Riemann tensor
\begin{equation}\label{Rie_Matrix}
    \begin{split}
        A\colon & \Lambda^+ \to \Lambda^+, \\
        \text{Riem} = \begin{pmatrix}
            A & B \\
            B^T & C
        \end{pmatrix} \quad \quad \quad B\colon & \Lambda^+ \to \Lambda^-, \\
        B^T\colon & \Lambda^- \to \Lambda^+, \\
        C \colon & \Lambda^- \to \Lambda^-.
    \end{split}
\end{equation}
A number of important questions in differential geometry and GR in four dimensions are a consequence of understanding how the matrix representation (\ref{Rie_Matrix}) is connected to the Ricci decomposition (\ref{Ricci_Decomp}). For instance, it can be shown that the off-diagonal components $B$ (the $3\times 3$ sector causing a mixing between Hodge eigenspaces) encodes the traceless Ricci tensor \cite{Besse1987, Krasnov2020}, i.e. 
\begin{equation}
    B = S_0^2(T^*\mathcal{M}) \KN g
\end{equation}
in the sense that $B$ vanishes \textit{if and only if} the trace-free Ricci tensor vanishes:
\begin{equation}\label{TF_Ricci}
    B = 0 \quad \Leftrightarrow \quad Z_{\mu \nu} = R_{\mu \nu} - \frac{1}{4} R g_{\mu \nu} = 0.
\end{equation}
It follows immediately that such a spacetime must be \textit{Einstein},
\begin{equation}
    R_{\mu \nu} \propto g_{\mu \nu},
\end{equation}
with proportionality fixed to be constant via the second (contracted) Bianchi identity. The blocks $A$ and $C$ (which are complex conjugates) encode the self-dual and anti-self-dual components of the Weyl tensor $W^{\pm}$ (respectively), together with the scalar curvature: this is the content of equation (\ref{Irr_Rep_Rie}). Spacetimes for which both $B = 0$ and $W^- = 0$ are the so-called \textit{self-dual} Einstein spacetimes, with examples including Eguchi–Hanson and Taub–NUT \cite{EguchiGilkeyHanson1980}. These are geometries for which the Twistor construction is naturally integrable \cite{Penrose1976NonlinearGravitons, MasonWoodhouse1996}. The characterization of the Riemann matrix in this form (in the sense of normal forms) is the origin of the Petrov classification \cite{Besse1987, Plebanski1977Separation, Petrov1969, Petrov2000Classification}.

The condition $B = 0$ can be represented instead as an operator identity on the space of 2-forms. Since in the basis (\ref{SD_Decomp}) the Hodge operator (\ref{Hodge_Op}) is diagonal
\begin{equation}
    \star = i
    \begin{pmatrix}
        \mathbb{I}_{3\times 3} & 0 \\
        0 & -\mathbb{I}_{3\times3}
    \end{pmatrix}
\end{equation}
with $\mathbb{I}_{3\times 3}$ the identity matrix on $\Lambda^{\pm}$, then the commutator
\begin{equation}
    \text{Riem} \;\star - \star \; \text{Riem} = [\text{Riem}, \star],
\end{equation}
or equivalently in tensor notation
\begin{equation}\label{Comm_Tensor}
    \frac{1}{2}R_{\mu \nu}^{\quad \rho \sigma} \varepsilon_{\rho \sigma }^{\quad \alpha \beta} - \frac{1}{2}\varepsilon_{\mu \nu}^{\quad \rho \sigma} R_{\rho \sigma}^{\quad \alpha \beta}, 
\end{equation}
extracts only of the off-diagonal components of the representation (\ref{Rie_Matrix}). This corresponds to the trace-free Ricci tensor (\ref{TF_Ricci}), and so a spacetime is Einstein \textit{if and only if} this commutator vanishes \cite{Besse1987, Krasnov2020}
\begin{equation}\label{EC_Formal}
    R_{\mu \nu}^{\quad \rho \sigma} \varepsilon_{\rho \sigma }^{\quad \alpha \beta} - \varepsilon_{\mu \nu}^{\quad \rho \sigma} R_{\rho \sigma}^{\quad \alpha \beta} = 0 \quad \Leftrightarrow \quad R_{\mu \nu} \propto g_{\mu \nu}.
\end{equation}

The above condition is particularly interesting in the context of GR. The field equations in vacuum 
\begin{equation}\label{EFE_Sec_2}
    R_{\mu \nu} - \frac{1}{2} R g_{\mu \nu} + \Lambda g_{\mu \nu} = 0,
\end{equation}
automatically enforce the Einstein condition 
\begin{equation}
    R_{\mu \nu} = \Lambda g_{\mu \nu}. 
\end{equation}
Thus, any vacuum spacetime in four dimensions automatically satisfies
\begin{equation}\label{Comm_Sec_2}
    [\text{Riem}, \star] = 0.
\end{equation}
Whether one works exclusively with the field equations (\ref{EFE_Sec_2}) or the commutator condition (\ref{Comm_Sec_2}) is equivalent \textit{up to the interpretation of $\Lambda$} \cite{Hughes2024}, as will be explained below. If the spacetime instead contains energy momentum $T_{\mu \nu}$, such that the full field equations (setting $c = G = 1$) are instead satisfied
\begin{equation}
    R_{\mu \nu} - \frac{1}{2} R g_{\mu \nu} + \Lambda g_{\mu \nu} = 8\pi T_{\mu \nu},
\end{equation}
then the commutator (\ref{Comm_Sec_2}) fails to close. Instead, (\ref{Comm_Tensor}) now serves as a characterization for the degree to which the spacetime fails to be Einstein. Stated differently, the operator 
\begin{equation}\label{delta_op}
    \Delta_{\mu \nu}^{\quad \alpha \beta} = \frac{1}{2} R_{\mu \nu}^{\quad \rho \sigma} \varepsilon_{\rho \sigma }^{\quad \alpha \beta} -  \frac{1}{2} \varepsilon_{\mu \nu}^{\quad \rho \sigma} R_{\rho \sigma}^{\quad \alpha \beta} 
\end{equation}
encodes at the level of $\mathcal{K}$ matters influence on the geometry of spacetime. Specifically non-zero energy-momentum activates the $\Lambda^+ \otimes \Lambda^-$ (self-dual anti-self-dual) sector of the geometry which the operator (\ref{delta_op}) is sensitive to. 

In what follows, we are interested in constructing the analogue of (\ref{Comm_Sec_2}) when energy-momentum does not vanish. In other words, we seek a lift of the field equations for gravitation to (a subset of) the algebraic space of curvature tensors $\mathcal{K}$. It turns out that this can be done by utilizing the KN product (\ref{KN_Prod}), but only up to the trace-free sector of the geometry. Provided then that one assumes the conservation of energy-momentum (as in the unimodular theory), this gives an equivalent reformulation of GR. We will then consider this reformulation in the electrovacuum sector.

\section{Trace-Free Field Equations as an Algebraic Condition}
When one extends from vacuum to non-empty spacetimes, the energy-momentum tensor $T_{\mu \nu}$ activates and the trace-free Ricci tensor becomes non-zero. As we have seen, in four spacetime dimensions $Z_{\mu \nu}$ corresponds to the $(1,1)$ irreducible representation of the Riemann tensor, which in the self-dual anti-self-dual basis (\ref{Rie_Matrix}) is extracted by forming the commutator with the Hodge dual (\ref{Delta}). Demanding that this commutator vanish is algebraically equivalent to imposing the vacuum field equations, realized as the elimination of the $(1,1)$ the trace-free sector. In non-empty spacetimes, this condition no longer holds:
\begin{equation}
    T_{\mu \nu} \neq 0 \implies Z_{\mu \nu} \neq 0  \quad \Leftrightarrow \quad [\text{Riem}, \star] \neq 0.
\end{equation}
The question we are concerned with is how this commutator condition can be generalized in terms of non-vanishing energy-momentum tensor. That is, can we formally represent the role of non-vacuum as a constraint on the algebraic space of curvature tensors $\mathcal{K}$? Below it will be shown, via the KN product, that imposing the trace-free field equations is equivalent to imposing the following condition on $\mathcal{K}$,
\begin{equation}\label{Main_Equations}
    [\text{Riem}, \star] = 4\pi [T \KN g, \star].
\end{equation}
Prior to this, it is worth recalling the conceptual differences between Einstein's theory and the unimodular theory (or what is effectively the trace-free theory), which for all intents and purposes are observationally indistinguishable \cite{Ellis2014TraceFreeInflation, CarballoRubioGarayGarciaMoreno2023, BengocheaLeonPerezSudarsky2023}.

\subsection{Unimodular Gravity}
The Einstein field equations (\ref{EFE_Sec_2}) contain two important pieces of information beyond their field content. The first is that through the contracted Bianchi identity, we determine the covariant conservation of energy-momentum:
\begin{equation}\label{Cont_Bianchi}
    \nabla_{\mu} \bigg(R^{\mu \nu} - \frac{1}{2}R g^{\mu \nu} + \Lambda g^{\mu \nu} \bigg) = 0 \quad \Rightarrow \quad \nabla_{\mu }T^{\mu \nu} = 0.
\end{equation}
One can alternatively deduce this conservation using the diffeomorphism invariance of the Einstein-Hilbert action with a minimally coupled matter Lagrangian \cite{Carroll1997GRNotes}. 

The second point is that the Einstein field equations (\ref{EFE_Sec_2}) determine the trace sector of the geometry through contraction with the inverse metric $g^{\mu \nu}$:
\begin{equation}\label{Trace_Sector}
    R  + 8\pi T = 4\Lambda.
\end{equation}

If one instead works with only the trace-free field equations,
\begin{equation}\label{TF_FE}
    R_{\mu \nu} - \frac{1}{4} R g_{\mu \nu} = 8\pi \bigg(T_{\mu \nu} - \frac{1}{4} T g_{\mu \nu}\bigg)
\end{equation}
then \textit{both} of the above pieces of information are lost. Instead one can take the covariant derivative of (\ref{TF_FE}) and use the contracted Bianchi identity (\ref{Cont_Bianchi}) to yield 
\begin{equation}
    g^{\mu \nu} \nabla_{\mu} R = 32\pi \nabla_{\mu}T^{\mu \nu} - 8\pi g^{\mu \nu} \nabla_{\mu} T. 
\end{equation}
Now, if it is \textit{assumed} that the covariant derivative of the energy-momentum is conserved (see \cite{BengocheaLeonPerezSudarsky2023} for a technical discussion surrounding this point), then the above reduces to
\begin{equation}\label{Key_Assumption}
    \nabla_{\mu} \big( R + 8\pi T) = 0 \quad \Rightarrow \quad R + 8\pi T = 4\Lambda,
\end{equation}
where unlike in (\ref{Trace_Sector}), $4\Lambda$ is now understood as an integration constant (which has strong implications for the cosmological constant problem \cite{Ellis2014TraceFreeInflation, Kaloper2014_2, Weinberg1989}). Therefore, while Einstein's GR requires only the field equations (\ref{EFE_Sec_2}) - or the associated action principle - trace-free gravity requires the vanishing $\nabla_{\mu} T^{\mu \nu}$ by hand \cite{Ellis2014TraceFreeInflation}. 

\subsection{Reformulating Unimodular Gravity}
Suppose then that there exists $T_{\mu \nu} \in S^2(T^*\mathcal{M})$, such that the Einstein field equations are satisfied:
\begin{equation}
    R_{\mu \nu} - \frac{1}{2} R g_{\mu \nu} + \Lambda  g_{\mu \nu} = 8\pi T_{\mu \nu}. 
\end{equation}
The energy-momentum tensor has a natural lift to the algebraic space of curvature tensors $\mathcal{K}$ via the action of the KN product (\ref{KN_Prod}). In fact, we may lift directly to the space $\mathcal{P}_g$ (i.e. the \textit{non}-Weyl component, \ref{P_g}) via the KN product with the metric:
\begin{equation}
    \big(T \KN g\big)_{\mu \nu \rho \sigma} = T_{\mu \rho} g_{\nu \sigma} + T_{\nu \sigma} g_{\mu \rho} - T_{\mu \sigma} g_{\nu \rho} - T_{\nu \rho} g_{\mu \sigma},
\end{equation}
or raising the second pair of indices (hence viewing this tensor as an operator on the 2-forms), what can instead be written as 
\begin{equation}
    \big(T \KN g\big)_{\mu \nu}^{\quad \rho \sigma} = T_{\mu}^{\rho} \delta_{\nu}^{\sigma} + T_{\nu}^{\sigma} \delta_{\mu}^{\rho} - T_{\mu}^{ \sigma} \delta_{\nu}^{\rho} - T_{\nu}^{\rho} \delta_{\mu}^{\sigma}.
\end{equation}
This tensor field has the following important contractions:
\begin{equation}\label{Contractions}
    \tilde{T}^{\sigma}_{\nu} \coloneq \big(T \KN g\big)_{\mu \nu}^{\quad \mu \sigma} = T \delta_{\nu}^{\sigma} + 2T_{\nu}^{\sigma}, \quad \tilde{T}_{\sigma}^{\sigma} = \big(T \KN g\big)_{\mu \nu}^{\quad \mu \nu} = 6 T.
\end{equation}

The claim is that the trace-free field equations correspond to the condition
\begin{equation}
    [\text{Riem}, \star] = 4\pi [T \KN g, \star].
\end{equation}
To prove this, we write the commutator as follows:
\begin{equation}\label{Main_Eqn_Indices}
    \frac{1}{2} R_{\gamma \delta}^{\quad \alpha \beta} \varepsilon_{\alpha \beta}^{\quad \rho \sigma} -  \frac{1}{2}\varepsilon_{\gamma \delta}^{\quad \alpha \beta} R_{\alpha \beta}^{\quad \rho \sigma} = 4\pi \bigg[\frac{1}{2}\big(T \KN g\big)_{\gamma \delta}^{\quad \alpha \beta} \varepsilon_{\alpha \beta}^{\quad \rho \sigma} - \frac{1}{2}\varepsilon_{\gamma \delta}^{\quad \alpha \beta} \big(T \KN g\big)_{\alpha \beta}^{\quad \rho \sigma}\bigg]. 
\end{equation}
Taking the dual over $(\gamma \delta)$ and using the fact that the Hodge dual squares to minus the identity (\ref{Hodge_Square_Prop}), we see that
\begin{equation}\label{Step_1}
    \frac{1}{4}\varepsilon_{\mu \nu}^{\quad \gamma \delta } R_{\gamma \delta}^{\quad \alpha \beta} \varepsilon_{\alpha \beta}^{\quad \rho \sigma} + R_{\mu \nu}^{\quad \rho \sigma} = 4\pi \bigg[\frac{1}{4} \varepsilon_{\mu \nu}^{\quad \gamma \delta }\big(T \KN g\big)_{\gamma \delta}^{\quad \alpha \beta} \varepsilon_{\alpha \beta}^{\quad \rho \sigma} + \big(T \KN g\big)_{\mu \nu}^{\quad \rho \sigma} \bigg].
\end{equation}
Using now the standard identity of contraction in the Levi-Civita tensor
\begin{equation}
    \varepsilon_{\gamma \delta \mu \nu} \varepsilon^{\alpha \beta \rho \sigma} = -24 \delta_{[\gamma}^{\alpha} \delta_{\delta}^\beta \delta_{\mu}^{\rho} \delta_{\nu]}^{\sigma},
\end{equation}
one observes that for any tensor $M_{\mu \nu}^{\quad \rho \sigma}$ satisfying the symmetries of the Riemann tensor (\ref{Rie_Symm}) \cite{Krasnov2020},
\begin{equation}
    \frac{1}{4} \varepsilon_{\mu \nu}^{\quad \gamma \delta} M_{\gamma \delta}^{\quad \rho \sigma} \varepsilon_{\rho \sigma}^{\quad \alpha \beta} = -M_{\mu \nu}^{\quad \alpha \beta} - M \delta_{[\mu}^{\alpha} \delta_{\nu]}^{\beta} + 2 \delta^{\alpha}_{[\mu}M_{\nu]}^{\beta} - 2 \delta_{[\mu}^{\beta} M_{\nu]}^{\alpha},
\end{equation}
where 
\begin{equation}
    M^{\beta}_{\nu } \equiv M_{\alpha \nu}^{\quad \alpha \beta}, \quad M \equiv M^{\alpha}_{\alpha}.
\end{equation}
With this, (\ref{Step_1}) reduces to 
\begin{equation}\label{Step_2}
    -R \delta_{[\mu}^\rho \delta_{\nu]}^{\sigma} + 2 \delta_{[\mu}^{\rho} R_{\nu]}^{\sigma} - 2 \delta_{[\mu}^{\sigma} R_{\nu]}^{\rho} = 4\pi\big(-\tilde{T}_\alpha^{\alpha} \delta_{[\mu}^\rho \delta_{\nu]}^{\sigma} + 2 \delta_{[\mu}^{\rho} \tilde{T}_{\nu]}^{\;\sigma} - 2 \delta_{[\mu}^{\sigma} \tilde{T}_{\nu]}^{\;\rho}\big).
\end{equation}
This however is equivalent to the trace-free field equations. Indeed, contracting over $(\mu ,\rho)$ and reintroducing $T_{\mu \nu}$ via (\ref{Contractions}), we have
\begin{equation}
    2R_{\nu}^{\sigma} - \frac{1}{2} R \delta^{\sigma}_{\nu} = 4\pi \big(4 T_{\nu}^{\sigma} - T \delta_{\nu}^{\sigma}\big),
\end{equation}
or equivalently
\begin{equation}\label{Trace_Free_Field_Equations}
    R_{\mu \nu} - \frac{1}{4} R g_{\mu \nu} = 8\pi\big(T_{\mu \nu} - \frac{1}{4} T g_{\mu \nu}\big).
\end{equation}
Provided then that the condition $\nabla_{\mu} T^{\mu \nu} = 0$ is satisfied, equation (\ref{Main_Equations}) represents an equivalent reformulation of GR unique to four dimensions.  Furthermore, it is clear that the Einstein condition (\ref{EC_Formal}) is appropriately recovered in the limit $T_{\mu \nu} \to 0$.

\subsection{Energy-Momentum Conservation}
Just as we can recast the content of the field equations as a constraint on the space $\mathcal{K}$, we can achieve a similar statement regarding energy-momentum conservation, thus expressing GR (via equivalence to the unimodular theory with energy-momentum conservation) as a pair of statements on objects naturally defined in $\mathcal{K}$. As we have seen, the Hodge dual in four dimensions splits the space of 2-form into its (complex) eigenspaces (\ref{Hodge_Decomp}). Given any 2-form, we can form the duality projection operator $P_{\mu \nu}^{\pm \; \rho \sigma}$ into the associated self-dual and anti-self-dual eigenspaces via \cite{Krasnov2020}
\begin{equation}\label{Projetion_Operator}
    P_{\mu \nu}^{\pm \; \rho \sigma} = \frac{1}{2} \bigg(\delta^{\rho}_{[\mu} \delta^{\sigma}_{\nu]} \mp \frac{i}{2} \varepsilon_{\mu \nu}^{\quad \rho \sigma}\bigg).
\end{equation}
In order to isolate the trace elements of the curvature and energy momentum - as in (\ref{Key_Assumption}) - we need to isolate the diagonal components of (\ref{Rie_Matrix}) and remove the contribution from the Weyl tensor. Since $\mathcal{W}$ is tracefree, this is done simply as follows for the curvature
\begin{equation}
    \text{Tr}\big(P^{\pm} \; \text{Riem} \; P^{\pm} \big) = \text{Tr}\big(P^{\pm} \;\text{Riem}\big) = \frac{1}{2} \bigg(\delta^{\rho}_{[\mu} \delta^{\sigma}_{\nu]} \mp \frac{i}{2} \varepsilon_{\mu \nu}^{\quad \rho \sigma}\bigg) R_{\rho \sigma}^{\quad \mu \nu} = \frac{R}{2}. 
\end{equation}
Similarly, given an energy momentum tensor $T_{\mu \nu} \in S^2(T^*\mathcal{M})$, its lift into $\mathcal{K}$ via the KN product with the metric contains both trace-free and scalar irreducible representations (but \textit{not} a Weyl-like component, see \ref{Ricci_Decomp}) to give a block $6\times6$ matrix in the self-dual anti-self-dual basis (\ref{Rie_Matrix})
\begin{equation}
    T \KN g = \begin{pmatrix}
        \mathbb{R} g \KN g & g \KN T_0 \\
        [g \KN T_0]^T & \mathbb{R} g \KN g
    \end{pmatrix}. 
\end{equation}
Consequently using (\ref{Contractions}) we have 
\begin{equation}
    \text{Tr}\big(P^{\pm} \; T\KN g\big) = 3T.
\end{equation}
The conservation of energy-momentum - and its implication in the form of (\ref{Key_Assumption}) - therefore amounts to (assuming the field equations are already imposed) 
\begin{equation}\label{EM_Cons}
    \nabla_{\mu} \; \text{Tr}\bigg( P^{\pm}  \bigg[2\; \text{Riem}   +  \frac{8\pi}{3} g \KN T\bigg]\bigg) = 0 \quad \Leftrightarrow \quad \nabla_{\mu} T^{\mu \nu} = 0.
\end{equation}
Thus, taking conditions (\ref{Main_Equations}) and (\ref{EM_Cons}) together in four dimensions is equivalent to imposing Einstein's field equations.

\section{Example: Electrovacuum Solutions}
Let us consider some clarifying examples of how the commutator condition (\ref{Main_Equations}) can be solved prior to the projection to the trace-free field equations. Below we will consider Reissner-Nordstr\"om and show that the commutator is identically satisfied. Following this, we will take the spherically symmetric ansatz with connection $A = -Q/r \; dt$ and show the commutator can be solved exactly to yield Reissner-Nordstr\"om de Sitter, giving an interesting insight into the role of the cosmological constant $\Lambda$. 

Electrovacuum solutions \cite{StephaniKramerMacCallumHoenselaersHerlt2003, GriffithsPodolsky2009} are spacetimes that satisfy the Einstein field equations 
\begin{equation}\label{EFE_Vac}
    R_{\mu \nu} - \frac{1}{2} R g_{\mu \nu} + \Lambda g_{\mu \nu} = \kappa T_{\mu \nu},
\end{equation}
where the energy-momentum tensor $T_{\mu \nu}$ is built from the anti-symmetric Maxwell tensor $F_{\mu\nu} = -F_{\nu \mu}$ via
\begin{equation}\label{EM_Tensor}
    T_{\mu \nu} = \frac{1}{4\pi} \bigg(g^{\alpha \beta} F_{\alpha \mu} F_{\beta \nu} - \frac{1}{4}g_{\mu \nu} F_{\alpha \beta} F^{\alpha \beta} \bigg).
\end{equation}
The electromagnetic field must also satisfy its associated covariant equations of motion, namely the Maxwell equations
\begin{equation}
    \nabla_{\mu} F^{\mu \nu} = 0
\end{equation}
together with the Bianchi identity in $F$. Since this system is manifestly not vacuum (i.e. $T_{\mu \nu} \neq 0$), then the Einstein condition (\ref{EC_Spacetime}) is not satisfied, $R_{\mu \nu} \neq \Lambda g_{\mu \nu}$. In other words,  the commutator between the Riemann tensor and the Hodge dual in electrovacuum spacetimes \textit{fails to close}:
\begin{equation}
    \Delta_{\mu \nu}^{\quad \alpha \beta} = \frac{1}{2}R_{\mu \nu}^{\quad \rho \sigma} \varepsilon_{\rho \sigma }^{\quad \alpha \beta} - \frac{1}{2}\varepsilon_{\mu \nu}^{\quad \rho \sigma} R_{\rho \sigma}^{\quad \alpha \beta} \neq 0.
\end{equation}
The (2,2)-tensor field $\Delta_{\mu \nu}^{\quad \alpha \beta}$ representing this commutator possesses the symmetries of the Riemann tensor (\ref{Rie_Symm}) \textit{and} satisfies the first Bianchi identity \cite{Weinberg1972}
\begin{equation}
    \varepsilon^{\mu \nu \rho \sigma} \Delta_{\mu \nu \rho \sigma} = 0.
\end{equation}
As we described in the previous section, the failure of the closure between the commutator of the Riemann tensor and the Hodge dual is characterized by the lift of the energy-momentum tensor (\ref{EM_Tensor}) to the algebraic space of curvature tensors $\mathcal{K}$ (\ref{Ricci_Decomp}). To see how this works in practice, consider the Reissner-Nordstr\"om spacetime
\begin{equation}\label{RN_Spacetime}
    ds^2 = -\bigg(1 - \frac{2M}{r} + \frac{Q^2}{r^2}\bigg)dt^2 + \frac{1}{1 - \frac{2M}{r} + \frac{Q^2}{r^2}} dr^2 + r^2 d\theta^2 + r^2 \sin^2{\theta} \:d\phi^2.
\end{equation}
 The metric is spherically symmetric with $F_{\mu \nu}$ the components of the electromagnetic 2-form $F$ generated by the connection $A$ for a spherical charge distribution of strength Q,
\begin{equation}\label{EM_Strength}
    A = -\frac{Q}{r}dt, \quad F = dA = -\frac{Q}{r^2} dt \wedge dr.
\end{equation} 
The components of $\Delta_{\mu \nu}^{\quad \alpha \beta}$ can be computed explicitly, and there are only two non-trivial (up to symmetry) components (either of which can be used to compute the other via index manipulations):
\begin{equation}
    \Delta_{tr}^{\quad \theta \phi} = -\frac{2Q^2}{r^6 \sin{\theta}}, \quad \Delta_{\theta \phi}^{\quad tr} = - \frac{2Q^2 \sin{\theta}}{r^2}.
\end{equation}
It is clear from this form that the $Q\to0$ limit closes the commutator and consequently results in the spacetime being Einstein (\ref{EC}): this coincides with the limit in which Reissner-Nordstr\"om collapses to Schwarzschild, as expected (\ref{EC_Formal}) \cite{Hughes2024}. From (\ref{EM_Strength}), the energy-momentum tensor has the following diagonal form after raising an index:
\begin{equation}
    T^{\mu}_{\nu} = \frac{Q^2}{8\pi r^4} \; \text{diag}(-1,-1,1,1).
\end{equation}
We now form the lift of the EM tensor to the space $\mathcal{K}$ (\ref{Ricci_Decomp}) using the KN product with the metric (\ref{KN_Prod}):
\begin{equation}
    (T \KN g)_{\mu \nu}^{\quad \rho \sigma} = T_{\mu}^{\rho} \delta_{\nu}^{\sigma} + T_{\nu}^{\sigma} \delta_{\mu}^{\rho} - T_{\mu}^{ \sigma} \delta_{\nu}^{\rho} - T_{\nu}^{\rho} \delta_{\mu}^{\sigma}.
\end{equation}
The result gives two non-trivial components up to symmetry, namely
\begin{equation}
    (T \KN g)_{tr}^{\quad tr} = - (T \KN g)_{\theta \phi}^{\quad \theta \phi} = -\frac{Q^2}{4\pi r^4}.
\end{equation}
Designating the commutator between $(T \KN g)_{\mu \nu}^{\quad \rho \sigma}$ and the Hodge dual as 
\begin{equation}
    \mathcal{A}_{\mu \nu}^{\quad \alpha \beta} = \frac{1}{2}(T \KN g)_{\mu \nu}^{\quad \rho \sigma} \varepsilon_{\rho \sigma }^{\quad \alpha \beta} - \frac{1}{2}\varepsilon_{\mu \nu}^{\quad \rho \sigma} (T \KN g)_{\rho \sigma}^{\quad \alpha \beta},
\end{equation}
this tensor is observed to have only two non-trivial components: 
\begin{equation}
    \mathcal{A}_{tr}^{\quad \theta \phi} = -\frac{Q^2}{2\pi r^6 \sin{\theta}}, \quad Q_{\theta \phi}^{\quad tr} = -\frac{Q^2 \sin{\theta}}{2\pi r^2}
\end{equation}
and we observe immediately that (\ref{Main_Equations}) is indeed satisfied,
\begin{equation}
    \Delta_{\mu \nu}^{\quad \rho \sigma} = 4\pi \mathcal{A}_{\mu \nu}^{\quad \rho \sigma}.
\end{equation}
Similar results can be demonstrated to hold in other exact electrovacuum spacetimes, such as the Berotti-Robinson and Melvin solutions \cite{GarfinkleGlass2011}. 

\subsection{Spherical Symmetry}
As opposed to the formulation of analysis of (\ref{Main_Equations}) in Reissner-Nordstr\"om spacetime, we can instead begin from the spherically symmetric ansatz \cite{Straumann2013GR}
\begin{equation}\label{ansatz}
    ds^2 = -e^{2v(t,r)}dt^2 + e^{2f(t,r)} dr^2 +  r^2 d\theta^2 + r^2 \sin^2{\theta} \:d\phi^2
\end{equation}
and look to derive Reissner-Nordstr\"om from (\ref{EM_Strength}) together with (\ref{Main_Equations}) as an algebraic identity to be solved. This turns out to be a simple exercise due to the structure of $\Delta_{\mu \nu}^{\quad \rho \sigma}$ in the spherically symmetric background. Indeed, from the ansatz (\ref{ansatz}) $\Delta_{\mu \nu}^{\quad \rho \sigma}$ has (up to symmetries and raising/lowering of indicies) three non-trivial components (see also \cite{Hughes2024}):
\begin{equation}\label{First_Conditions}
    \begin{split}
        \Delta^{\quad t\theta}_{t\phi} & = - \Delta^{\quad r\theta}_{r\phi} = \frac{2 e^{-(f+v)}\sin{\theta}}{r} \frac{\partial f}{\partial t}
        \\
        \Delta^{\quad t\theta}_{r\phi} & = -\sin^2{\theta}\;\Delta^{\quad t\phi}_{r \theta} = \frac{ e^{-(f + v)}\sin{\theta}}{r}\bigg(\frac{\partial f}{\partial r} + \frac{\partial v}{\partial r} \bigg) 
    \end{split} 
\end{equation}
and
\begin{equation}\label{Final_Cond}
    \begin{split}
        \Delta^{\quad tr}_{\theta \phi}=&\bigg(r^2 e^{2f}\frac{\partial^2 f}{\partial t^2} + r^2 e^{2f}\bigg(\frac{\partial f}{\partial t}\bigg)^2 - r^2 e^{2f}\frac{\partial f}{\partial t}\frac{\partial v}{\partial t} + r^2e^{2v} \frac{\partial f}{\partial r} \frac{\partial v}{\partial r} \\
        & - r^2e^{2v}\bigg(\frac{\partial v}{\partial r}\bigg)^2 - r^2e^{2v} \frac{\partial^2 v}{\partial r^2} - e^{2(f+v)} + e^{2v}\bigg) e^{-3(f + v)} \sin{\theta}.
    \end{split}
\end{equation}
We now set by construction the condition (\ref{Main_Equations})
\begin{equation}\label{Alg_Cond}
    \Delta_{\mu\nu}^{\quad \alpha \beta} = 4 \pi\mathcal{A}_{\mu \nu}^{\quad \alpha \beta},
\end{equation}
where
\begin{equation}
    \mathcal{A}_{\mu \nu}^{\quad \alpha \beta} = \frac{1}{2}(T \KN g)_{\mu \nu}^{\quad \rho \sigma} \varepsilon_{\rho \sigma }^{\quad \alpha \beta} - \frac{1}{2}\varepsilon_{\mu \nu}^{\quad \rho \sigma} (T \KN g)_{\rho \sigma}^{\quad \alpha \beta}.
\end{equation}
The tensor $\mathcal{A}_{\mu \nu}^{\quad \rho \sigma}$ has only a single non-trivial element up to symmetry:
\begin{equation}\label{Q_form}
    \mathcal{A}_{\theta \phi}^{\quad tr} = -\frac{Q^2 \sin{\theta}}{2\pi r^2} e^{-3(f +v)}
\end{equation}
Thus, enforcing (\ref{Alg_Cond}) results in three differential equations (up to symmetry) to be solved in $f(t,r)$ and $v(t,r)$. From (\ref{First_Conditions}), (\ref{Alg_Cond}) demands
\begin{equation}
    \begin{split}
        \frac{\partial f}{\partial t} = 0 \quad \text{and} \quad \frac{\partial f}{ \partial r} = - \frac{\partial v}{\partial r}.
    \end{split}
\end{equation}
Consequently we have $f = f(r)$ and (without loss of generality) $v = v(r) = -f(r)$. The final condition to be solved then is the matching between (\ref{Final_Cond}) and (\ref{Q_form}). This reduces to
\begin{equation}
    r^2 \bigg(\frac{d^2 f}{d r^2}\bigg) + 2r^2\bigg(\frac{d f}{d r}\bigg)^2 - \frac{2Q^2}{r^2} + e^{-2f} - 1 = 0.
\end{equation}
This equation can be solved exactly for $f$:
\begin{equation}\label{Final_Exact_Sol}
    f(r) = \frac{1}{2} \ln \bigg(1 - \frac{2c_1}{r} + \frac{Q^2}{r^2}  - 2c_2 r^2\bigg),
\end{equation}
where $c_1$ and $c_2$ are constants to be identified with the mass $M$ and cosmological constant $\Lambda$ respectively.

That the cosmological constant has appeared naturally in the solving of the algebraic condition (\ref{Alg_Cond}) is noteworthy. If one instead attempts to solve the Einstein field equations under the same conditions above, then whether one obtains Reissner-Nordstr\"om or Reissner-Nordstr\"om-de Sitter resolves to a \textit{choice}\footnote{Of course, the Lovelock theorem \cite{Lovelock1971, Lovelock1972} demands that $\Lambda g_{\mu \nu}$ be present in the most general theory of Einstein's gravity. However, the point is that one cannot get Reissner-Nordstr\"om without $\Lambda$ from solving (\ref{Alg_Cond}), while one can preemptively solve the field equations (\ref{EFE_Vac}) in the absence of $\Lambda$. (\ref{Alg_Cond}) is sensitive to the most general case.} in the presence of the term $\Lambda g_{\mu \nu}$. For (\ref{Alg_Cond}) however one does not implement $\Lambda$ ``by hand" in this way: its existence ($\Lambda \in \mathbb{R}$) is guaranteed by the algebraic structure one enforces on $\mathcal{K}$. The results of section 3 however make this point easily understood. Since the commutator condition (\ref{Main_Equations}) on $\mathcal{K}$ encodes the trace-free field equations, $\Lambda$ is guaranteed to arise as an \textit{integral of motion} through the constraint on $\mathcal{K}$ via (\ref{Key_Assumption}).

\subsection{Removal of the Energy-Momentum Tensor}
Four-dimensional electro-vacuum is a somewhat unique case in the sense that it is not technically necessary to introduce the energy-momentum tensor to generate the trace-free equations from $\mathcal{K}$. This is a consequence of the fact that the KN product (\ref{KN_Prod}), while defined on the symmetric 2-tensors, lifts a pair of 2-forms to $\mathcal{K}$ in an analogous way. In fact, as presented in (\ref{KN_Prod}) the KN product is a restriction of a more general product structure over the graded algebra formed by symmetric products of $p$-forms:
\begin{equation}
    \sum_{p = 0}^n S^2(\Lambda^p \; T^* \mathcal{M}).
\end{equation}
Regardless, we see that for two 2-forms the KN products maps them into $\mathcal{K}$ via
\begin{equation}\label{KN_Prod_2}
    \begin{split}
        \KN\colon \Lambda^2(T^*\mathcal{M}) \otimes \Lambda^2(T^*\mathcal{M}) & \to \mathcal{K}, \\
        \omega_{\mu \nu},\tau_{\mu \nu}\in \Lambda^2(T^*\mathcal{M}) & \mapsto (\omega \KN \tau)_{\mu \nu \rho \sigma} = \omega_{\mu \rho} \tau_{\nu \sigma} + \omega_{\nu \sigma} \tau_{\mu \rho} - \omega_{\mu \sigma} \tau_{\nu \rho} - \omega_{\nu \rho} \tau_{\mu \sigma}.
    \end{split}
\end{equation}
As opposed to lifting the energy-momentum tensor via the metric then, we may simply introduce the electromagnetic 2-form $F$ via the connection $F = dA$ and lift instead $F$ directly with itself (since the energy-momentum tensor is quadratic in $F$):
\begin{equation}
    \begin{split}
        (F \KN F)_{\mu \nu \rho \sigma} & = F_{\mu \rho} F_{\nu \sigma} + F_{\nu \sigma} F_{\mu \rho} - F_{\mu \sigma} F_{\nu \rho} - F_{\nu \rho} F_{\mu \sigma} \\
        & = 2(F_{\mu \rho} F_{\nu \sigma} - F_{\mu \sigma} F_{\nu \rho})
    \end{split}
\end{equation}
The product $(F \KN F)_{\mu \nu \rho \sigma}$ has the following important contractions
\begin{equation}
    \tilde{F}_{\nu}^{\;\beta} =(F \KN F)_{\mu \nu}^{ \quad \mu \sigma}=  -2 F_{\nu}^{\;\alpha} F_{\alpha}^{\;\beta}, \quad \tilde{F}^{\;\alpha}_{ \alpha} = (F \KN F)_{\mu \nu}^{ \quad \mu \nu} = -2 F_{\beta}^{\;\alpha} F_{\alpha}^{\;\beta}.
\end{equation}
Following the derivation of section 3, it is straight-forward to verify that the electrovacuum field equations are recovered from setting 
\begin{equation}
    [\text{Riem}, \star] = [F \KN F, \star], \quad \Leftrightarrow \quad R_{\mu \nu} - \frac{1}{4}R g_{\mu \nu} = 2\bigg(g^{\alpha \beta} F_{\alpha \mu} F_{\beta \nu} - \frac{1}{4}g_{\mu \nu} F_{\alpha \beta} F^{\alpha \beta} \bigg).
\end{equation}
That is to say in electrovaccum we have the identity on $\mathcal{K}$
\begin{equation}
    4\pi [T\KN g, \star] = [F \KN F, \star]
\end{equation}
with the energy-momentum tensor defined by (\ref{EM_Tensor}). Of course, this relation will hold for arbitrary gauge group $G$ where the appropriate trace is taken over gauge indices in the definition of energy-momentum. 

\section{Conclusions}
In this paper we were concerned with the unimodular (trace-free) theory in four dimensions \cite{Ellis2014TraceFreeInflation, CarballoRubioGarayGarciaMoreno2023, BengocheaLeonPerezSudarsky2023}, understood as a framework sharing the same moduli space as GR (hence an equivalent reformulation) but with the status of energy-momentum conservation and the cosmological constant altered. Four dimensions is interesting in differential geometry because the orthogonal Ricci decomposition of the Riemann tensor admits an irreducible representation over the self-dual anti-self-dual basis of the space $S^2 (\Lambda^2 \; T^*\mathcal{M})$ \cite{Besse1987}. In the algebraic space of curvature tensors $\mathcal{K}$, the trace-free sector appears in a natural way and forms a space of homomorphism operators on the 2-forms $\Lambda^2$
\begin{equation}
    \text{Hom}(\Lambda^+, \Lambda^-) \cong (1,0) \otimes (0,1) \cong (1,1).
\end{equation}
Since the Hodge dual is diagonal in this basis, the trace-free sector of $\mathcal{K}$ can always be extracted by forming the commutator with the Hodge dual. The space $\mathcal{K}$ is of central importance in gravitation and by developing a representation theory orientated approach to spacetime, one sees immediately exactly what the field equations of a gravitational theory demand upon elements of the space $\mathcal{K}$. Empty spacetime is equivalent to the vanishing of the $(1,1)$ irreducible representation (\ref{Ricci_Decomp}), or equivalently the vanishing of the off-diagonal components of the Riemann tensors $6\times 6$ representation over the eigenspaces of the Hodge dual on 2-forms (\ref{Rie_Matrix}) \cite{Krasnov2020}. This can then be simply repackaged as the condition that Riemann tensor commute with the Hodge dual as endomorphism operators on $\Lambda^2$ (\ref{Comm_Sec_2}). 

The introduction of energy-momentum into spacetime can be understood as a smooth shifting away from the vacuum condition, in the sense that the commutator between the Riemann tensor and the Hodge dual now fails to close precisely because the element $g \KN S^2_0(T^*\mathcal{M})$ is now \textit{sourced} by $T_{\mu \nu}$. Lifting $T_{\mu \nu}$ to the algebraic space $\mathcal{K}$ via the KN product (\ref{KN_Prod}) consequently gives exactly this source component, which can be elegantly matched to the trace-free component of the Riemann tensor via conditions on the commutator with the Hodge dual. Consequently, instead of describing this interaction at the level of the field equations, one can simply regard the field equations as a projection down from the appropriate conditions on $\mathcal{K}$ itself (\ref{Main_Equations}). It is clearly not possible to get Einstein's equations directly in this way, but only the trace-free sector of the geometry. However, if we couple the constraint on $\mathcal{K}$ with energy-momentum conservation (\ref{Key_Assumption}), one has the entire dynamical structure of GR. Hence, GR can be expressed in four dimensions only as the following pair of conditions on $\mathcal{K}$:
\begin{equation}\label{Main_Theory}
    \begin{split}
        [\text{Riem}, \star] = 4\pi [T \KN g, \star] \quad & \Leftrightarrow \quad R_{\mu \nu} - \frac{1}{4} R g_{\mu \nu} = 8\pi \bigg( T_{\mu \nu} - \frac{1}{4} T g_{\mu \nu}\bigg) \\
        \nabla_{\mu} \; \text{Tr}\bigg( P^{\pm}  \bigg[2\; \text{Riem}   +  \frac{8\pi}{3} g \KN T\bigg]\bigg) = 0 \quad & \Leftrightarrow \quad \nabla_{\mu} T^{\mu \nu} = 0.
    \end{split}
\end{equation}
In section 4 it was demonstrated that while this condition holds identically on (for instance) Reissner-Nordstr\"om, the commutator condition (\ref{Main_Equations}) can be solved without projecting in the case of charged spherical symmetry to yield Riessner-Nordstr\"om de Sitter uniquely. Just as unimodular gravity recovers the cosmological constant as a constant of motion \cite{Ellis2014TraceFreeInflation, CarballoRubioGarayGarciaMoreno2023, BengocheaLeonPerezSudarsky2023}, one sees explicitly that upon solving (\ref{Main_Theory}) $\Lambda$ arises in an equivalent manner, verifying the generality of the condition (\ref{Main_Equations}) on $\mathcal{K}$. 

The demonstration that a theory admits equivalent reformulations is an interesting observation, since it typically highlights new spaces or variables of relevance in which to frame its dynamical content. For GR proper there is an abundance of such reformulations typically centered on the soldering of various vector spaces (in the sense of Cartan), together with Lagrangian formalisms yielding the field equations on variation of the solder-form (e.g. Einstein-Cartan and chiral-BF theories) \cite{Krasnov2020, Krasnov2011, Mielke2017, BlagojevicHehl1995}. We have demonstrated here an analogous restructuring for the unimodular theory, in which its foundations are demonstrated to lie in representation theory as opposed to a strict action principle. This opens up the possibility of building more transparent contrasts between gravity and gauge theory, since a representation theory first perspective (at the cost of reducing from Einstein's theory to unimodular gravity, hence a less straightforward action principle \cite{BengocheaLeonPerezSudarsky2023}) adapted to the Hodge decomposition (\ref{Irr_Rep_Rie}) is exactly the starting point for the Plebanski formalism and the relationship of GR to the Ashtekhar variables and Twistor theory, where the addition of energy-momentum content is often highly non-trivial (while here it is a specific kind of operator in an irreducible representation) \cite{Plebanski1977Separation, Krasnov2017SelfDualGravity, NakamichiSugamotoOda1991, CapovillaDellJacobsonMason1991}. In the concrete sense demonstrated in this paper, energy-momentum can be regarded as an operator embedded in the self-dual anti-self-dual sector of the geometry (\ref{Irr_Rep_Rie}) which in vacuum is otherwise absent (\ref{TF_Ricci}): this is the space of relevance as far as the coupling between gravity and matter is concerned. Mapping (\ref{Main_Theory}) to the Plebanski soldering forms and contrasting the unimodular theory to the various reformulations of GR is consequently an interesting direction for future research. It is also interesting to ask how the `inverse' problem of using (\ref{Main_Theory}) to determine $F$ for a given spacetime overlaps with Rainich conditions that motivated the geometrodynamics program \cite{Mielke2017, Rainich1925ElectrodynamicsGR, MisnerWheeler1957}.

\section{Acknowledgements}
Jack. C. M. Hughes is grateful to Joudy Jamal Beek and Zihao Chen for valuable discussions.

\section{Funding}
This work received institutional funding from Khalifa University. Jack C. M. Hughes and Fedor V. Kusmartsev acknowledge support from the Khalifa University Award Numbers FSU-2021-030/8474000371 as well as Research and
Innovation Grants RIG-2023-028, RIG-2024-26, and
RIG-2024-053. Fedor V. Kusmartsev also acknowledges
support from the Thousand Talents Program and the
President’s International Fellowship Initiative of the
Chinese Academy of Sciences Awards.

\section{Conflict of Interest}
The authors of this work declare that they have no conflicts of interest.

\section{References}
\bibliographystyle{iopart-num}
\bibliography{bib_main_fix.bib}

\end{document}